\documentclass[10pt, conference, compsocconf]{IEEEtran}

\usepackage{amsmath}
\usepackage{amssymb}
\usepackage{longtable}
\usepackage{cite}
\usepackage{graphicx}
\usepackage{epstopdf}
\usepackage{dcolumn}
\usepackage{stfloats}

\usepackage{color} 
\begin{document}

\title{Scaling of city attractiveness for foreign visitors through big data of human economical and social media activity}

%\author{\IEEEauthorblockN{Stanislav Sobolevsky}
%\IEEEauthorblockA{SENSEable City Lab, MIT\\
%Cambridge, MA, USA\\
%Email: stanly@mit.edu}
%\and
%\IEEEauthorblockN{Iva Bojic}
%\IEEEauthorblockA{SENSEable City Lab, MIT\\
%Cambridge, MA, USA\\
%Email: ivabojic@mit.edu}
%\and
%\IEEEauthorblockN{Alexander Belyi}
%\IEEEauthorblockA{SENSEable City Lab, MIT\\
%Cambridge, MA, USA\\
%Email: abely@mit.edu}
%\and
%\IEEEauthorblockN{Izabela Sitko}
%\IEEEauthorblockA{Department of Geoinformatics - Z\_GIS, \\University of Salzburg\\
%Salzburg, Austria\\
%Email: izabela.sitko@sbg.ac.at}
%\and
%\IEEEauthorblockN{Bartosz Hawelka}
%\IEEEauthorblockA{Department of Geoinformatics - Z\_GIS, \\University of Salzburg\\
%Salzburg, Austria\\
%Email: bartosz.hawelka@sbg.ac.at}
%\and
%\IEEEauthorblockN{Juan Murillo Arias}
%\IEEEauthorblockA{New Technologies, BBVA\\
%Madrid, Spain\\
%Email: juan.murillo.arias@bbva.com}
%\and
%\IEEEauthorblockN{Carlo Ratti}
%\IEEEauthorblockA{SENSEable City Lab, MIT\\
%Cambridge, MA, USA\\
%Email: ratti@mit.edu}
%}

%\author{
%\IEEEauthorblockN{Stanislav Sobolevsky, Iva Bojic, Alexander Belyi, Carlo Ratti}
%\IEEEauthorblockA{SENSEable City Lab, MIT\\
%Cambridge, MA, USA\\
%Email: stanly@mit.edu, ivabojic@mit.edu, abely@mit.edu, ratti@mit.edu}
%\\
%\IEEEauthorblockN{Izabela Sitko, Bartosz Hawelka}
%\IEEEauthorblockA{University of Salzburg \\
%Salzburg, Austria\\
%Email: izabela.sitko@sbg.ac.at, bartosz.hawelka@sbg.ac.at} %I corrected the email of Bartosz
%\\
%\IEEEauthorblockN{Juan Murillo Arias}
%\IEEEauthorblockA{New Technologies, BBVA\\
%Madrid, Spain\\
%Email: juan.murillo.arias@bbva.com}
%}

\author{
\IEEEauthorblockN{Stanislav Sobolevsky, Iva Bojic, Alexander Belyi}
\IEEEauthorblockA{SENSEable City Lab, MIT\\
Cambridge, MA, USA\\
Email: stanly@mit.edu, ivabojic@mit.edu, abely@mit.edu}
\\
\IEEEauthorblockN{Juan Murillo Arias}
\IEEEauthorblockA{New Technologies, BBVA\\
Madrid, Spain\\
Email: juan.murillo.arias@bbva.com}
\and
\IEEEauthorblockN{Izabela Sitko, Bartosz Hawelka}
\IEEEauthorblockA{University of Salzburg \\
Salzburg, Austria\\
Email: izabela.sitko@sbg.ac.at, bartosz.hawelka@sbg.ac.at} %I corrected the email of Bartosz
\\
\IEEEauthorblockN{Carlo Ratti}
\IEEEauthorblockA{SENSEable City Lab, MIT\\
Cambridge, MA, USA\\
Email: ratti@mit.edu}
}

\maketitle

\begin{abstract}
Scientific studies investigating laws and regularities of human behavior are nowadays increasingly relying on the wealth of widely available digital information produced by human social activity. In this paper we leverage big data created by three different aspects of human activity (i.e., bank card transactions, geotagged photographs and tweets) in Spain for quantifying city attractiveness for the foreign visitors. An important finding of this papers is a strong superlinear scaling of city attractiveness with its population size. The observed scaling exponent stays nearly the same for different ways of defining cities and for different data sources, emphasizing the robustness of our finding. Temporal variation of the scaling exponent is also considered in order to reveal seasonal patterns in the attractiveness. %a linear assumption for the number of points of attraction within a city being proportional to the city size.

\end{abstract}

\begin{IEEEkeywords}
human mobility; big data; urban attraction; Flickr; Twitter; bank cards; tourism; superlinear scaling; discrete-choice model.

\end{IEEEkeywords}

\IEEEpeerreviewmaketitle

\section{Introduction}
\label{sec:introduction}

Every day people around the world leave more and more of their digital traces at places they visit. There is an impressive amount of papers leveraging such data for studying human behavior, including mobile phone records \cite{ratti2006mlu, calabrese2006real, girardin2008digital, quercia2010rse}, vehicle GPS traces \cite{santi2013taxi, kang2013exploring}, smart cards usage \cite{bagchi2005, lathia2012}, social media posts \cite{java2007we, szell2013, frank2013happiness} and bank card transactions \cite{sobolevsky2014mining, sobolevsky2014money}. Results of the studies could be applied for addressing to a wide range of policy and decision-making challenges, such as regional delineation \cite{ratti2010redrawing, sobolevsky2013delineating} or land use classification \cite{pei2014new, grauwin2014towards} for instance. Many works focus specifically on studying human mobility at urban \cite{gonzalez2008uih, kung2014exploring, hoteit2014estimating}, country \cite{amini2014impact} or even global scale \cite{hawelka2014,	paldino2015flickr}.

In this paper we consider international human mobility on the example of Spain by quantifying and analyzing the ability of different cities to attract foreign visitors (for various purposes) and proposing the mechanism that can explain the observed pattern. For the purpose of the study we use information about interactions between people and businesses registered through bank card transaction records and between people and urban spaces using geotagged photographs and tweets. By utilizing Flickr and Twitter datasets, as well as bank card transaction dataset provided by Banco Bilbao Vizcaya Argentaria (BBVA) bank, our goal is to investigate how city ability to attract foreign visitors depends on the city size. City attractiveness is defined as the absolute number of photographs, tweets or economical transactions made in the city by the foreign visitors. 

Of course this overall measure by itself does not tell the whole story as for example it can not explain the reason for the attractiveness --- is the city really a touristic destination or does it attract a lot of business or a more special category of visitors? And what actually makes it attractive? Also one might want to zoom into much more details here analyzing where those visitors come from and what specific places across the city they visit. Obviously this kind of analysis would also require an in-depth consideration of the city historical, cultural, demographical, economical and infrastructural context. While in this paper we'll focus just on a general overall attractiveness estimate as an initial step in this direction.

Discovering how to improve city attractiveness, which is seen differently by residents and tourists, can be used in several fields such as planning, forecasting flows, tourism, economics and transportation \cite{sinkiene2014concept}. In the past, photography was also already considered as a good mean of inquiry in architecture and urban planning, being used for understanding landscapes \cite{spirn1998language}. Girardin et al.\ showed that it was possible to define a measure of city attractiveness by exploring big data from photo sharing websites \cite{girardin2009quantifying}. Moreover, shopping patterns of tourists, including their specific preferences and satisfaction level, were analyzed with the overall purpose of accurate planning, marketing and management of sales strategies \cite{lehto2004, oh2004, yuksel2004, rosenbaum2005, uncles1995}. However, there was little work done to show how city attractiveness can be quantified and explained from a many-sided perspective of diverse datasets created by different aspects of individual activity. 

The novelty of this paper is twofold: applying new multi-layered data (i.e., Flickr, Tweeter and bank card transactions) for quantifying urban attractiveness for the foreign visitors as well as detecting strong and robust pattern. More specifically, our study looks at the way how attractiveness of a city depends on its size. Validating the robustness of the findings we look at different ways of city definitions and at different datasets used to quantify the attractiveness. 

\section{Datasets}
\label{data_sets}
In our analysis we are combining three different datasets: the first one containing more than 100 million publicly shared geotagged photographs on Flickr around the world took during a period of several years, the second containing geotagged tweets posted on Twitter worldwide during 2012, and the last one containing a set of bank card transactions of domestic and foreign users recorded by Banco Bilbao Vizcaya Argentaria (BBVA) during 2011, all over Spain. The aforementioned data allow us to analyze activity of the foreign tourists in Spain from three different aspects - making purchases, taking photographs or expressing sentiments of interesting places they visited.

\subsection{Flickr dataset}
By merging two Flickr datasets \cite{flickr1, flickr2} we created a new dataset containing more than 130 million photographs/videos. Both datasets are available upon request to the interested researchers~-- one coming from a research project, another from Yahoo. Each dataset consists of over 100 million photographs or videos taken by more than one million users. The records in two datasets partially overlap, but since each photograph/video in both datasets has its id, we were able to merge both datasets by omitting duplicates and choosing only those entries that were made within a 10 year time window, i.e., from 2005 and until 2014.
 
In order to determinate which of the users acting in a certain location are actually foreign visitors, for each user in the merged dataset we define his/her home country by using the following criteria: a person is considered to be a resident of a certain country if this is the country where he/she took the highest number of the photographs/videos over the longest timespan (calculated as the time between the first and last photograph taken within the country) compared to all other countries for the considered person. For over 500 thousand users we were able to determine their home country using our criteria. Those users took almost 80\% of all the photographs/videos in the dataset (i.e., more than 90 millions in total), while the rest of users for which home country can not be defined mostly belong to a low-activity group taking photographs only occasionally. 

For the purpose of our study we only consider the users with defined home country. From the total of over 3.5mln pictures taken in Spain, over 400 thousand are taken by over 16 thousand of foreign visitors coming from 112 countries all over the world.

\subsection{Twitter dataset}
The second dataset consists of geotagged messages posted during 2012 and collected from the digital microblogging and social media platform Twitter. Data was collected with the Twitter Streaming API \cite{twitterapi} and cleansed from potential errors and artificial tweeting noise as previously described in \cite{hawelka2014}. Globally, the dataset covers 944M tweets sent by 13M users \cite{hawelka2014}.
The final number of tweets posted in Spain in 2012 reached almost 35 million messages sent by 641 thousand Twitter users.
To differentiate between Spanish residents and foreign visitors, we used a similar technique as the one used in a case of Flickr dataset. We found out that 2\% of the total number of tweets posted in Spain in 2012 was sent by 80 thousand foreign visitors from 180 countries. 

\subsection{BBVA dataset}
The third dataset used in this study is a complete set of bank card transactions registered by the Spanish bank BBVA during 2011. Those transactions are of two types: i) made using debit or credit cards issued by BBVA, or ii) made using cards issued by other banks in any of over 300 thousand BBVA card terminals. For the transactions of the second group, the dataset includes the country of origin where the card was issued. For our study we focus mainly on this second group, in particular on 17 million transactions made by the 8.6 million foreigner visitors from 175 different countries.

Due to the sensitive nature of bank data, our dataset was anonymized by BBVA prior to sharing, in accordance to all local privacy protection laws and regulations. Therefore, cards are identified only by randomly generated IDs, and all the details that could allow re-identifying a card holder were removed. The raw dataset is protected by the appropriate non-disclosure agreement and is not publicly available. %However, the researchers may share certain aggregated data upon request for the purpose of validating the results of the conducted study.

\section{Scaling of city attractiveness in Spain}
\label{section_3}

Cities are known not only to be the places where people live, but also the environment transforming human life. A bigger city boosts up human activity: intensity of interactions \cite{schlapfer2012scaling}, creativity \cite{bettencourt2010urbscaling}, economic efficiency (e.g., measured in GDP \cite{bettencourt2013origins}), as well as certain negative aspects: crime \cite{bettencourt2010urbscaling} or infectious diseases \cite{bettencourt2007growth}. Due to agglomeration effects and intensified human interactions, many aggregated socioeconomic quantities are known to depend on the city size in the form of the superlinear scaling laws, meaning that those quantities are not simply growing with the city size, but are actually growing faster compared to it; at the other hand urban infrastructure dimensions (e.g., total road surface) reveal a sublinear relation to the city size \cite{batty2008size, bettencourt2007growth, brockmann2006scaling, bettencourt2013origins}. 

However, all of above mentioned quantities are mainly related to the processes that are happening inside the city. In this paper we pose another closely related, but slightly different question about such a characteristic of city external appearance as its attractiveness for the foreign visitors. Worth mentioning is that by "attractiveness" we do not necessary mean that a place is being a touristic destination, but we are just looking at its ability to attract the visitors for whatever reason~-- touristic, business, or any other personal matter-related ones.

In this study we focus on Spain as it is the country which economy largely depend on international tourism giving a paramount importance to the ability of its cities to attract foreign visitors (even if they are coming for a primary purpose other than tourism, visitors still often act like tourists do, making their contribution to the touristic sector). Many tourist rankings only consider a number of people visiting a city, consequently often leading to a fairly obvious conclusion that larger cities are more attractive as they can accommodate more tourists. The question we are interested in is how the total amount of visitor activity typically scales with the city size. Understanding this kind of scaling allows to predict the expected performance for a particular city and to estimate if it is actually under- or over-performing compared to the average expectation for the cities of that size.

The first step of conducting an analysis of city attractiveness is to get decided with what should be considered as a city. There is a number of ways of how to define a city, and selecting an appropriate city definition is important for analyzing aggregated urban performance \cite{batty2008size, roth2011structure, bettencourt2013origins}. For the purpose of our study we utilized definitions proposed by European Urban Audit Survey (EUAS) \cite{urbauditweb}, European Spatial Planning Observation Network (ESPON) \cite{espon2007} and A\'reas Urbanas de Espan\~a (AUDES) \cite{audes} project. On the most fine grain level, AUDES project defines 211 conurbations (CONs) in Spain. On more aggregate level, ESPON defines 40 Functional Urban Areas (FUAs), and finally on the most aggregated level EUAS defines 24 Large Urban Zones (LUZs). Population for LUZ and FUA were obtained from Eurostat \cite{eurostat} and National Statistics Institute of Spain \cite{ine}, and for CONs from the AUDES project. In addition to those three city definitions, we also perform our analysis for the 52 Spanish provinces in order to see if our conclusions could be actually extrapolated from the level of cities to more general consistent geographical entities.

\subsection{Aggregated city attractiveness}
\label{section_3.1}

In order to explore the overall city attractiveness for the foreign visitors we focus on three different aspects of visitor activity - economical transactions, taking photographs and twitting during their visit. We use the total amount of the described activity to quantify the attractiveness measure instead of a simple count of the users, since unlike the quantity of people who visited the city at least once, the amount of activity also takes into account the length of visitors' stay and intensity of exploring the city. We believe this is a more relevant proxy for the average visitor activity in the observed city at every single moment of time. 

\begin{figure*}[t!]
\centering
\includegraphics[width=.49\textwidth]{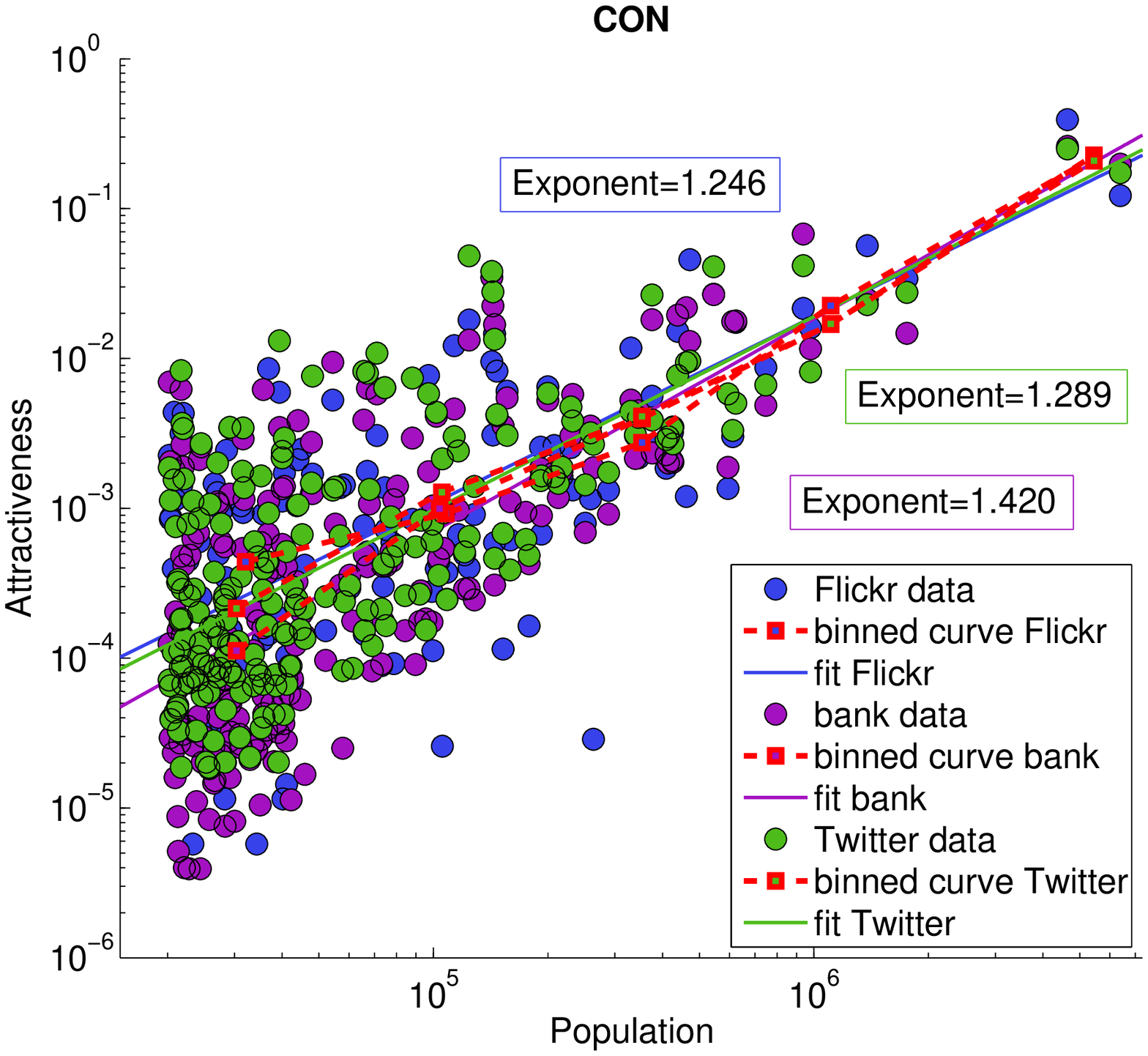}
\includegraphics[width=.49\textwidth]{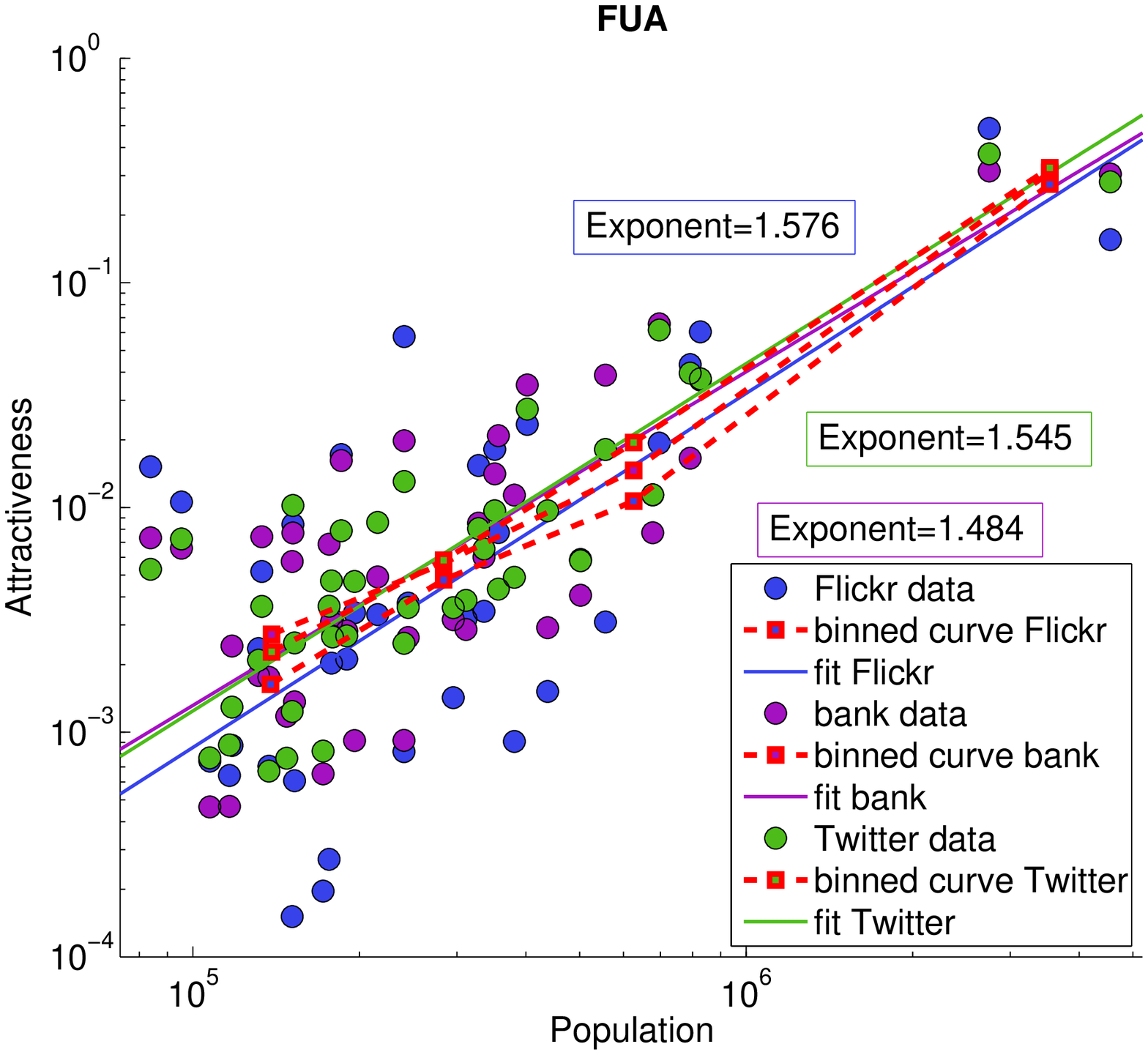}
\includegraphics[width=.49\textwidth]{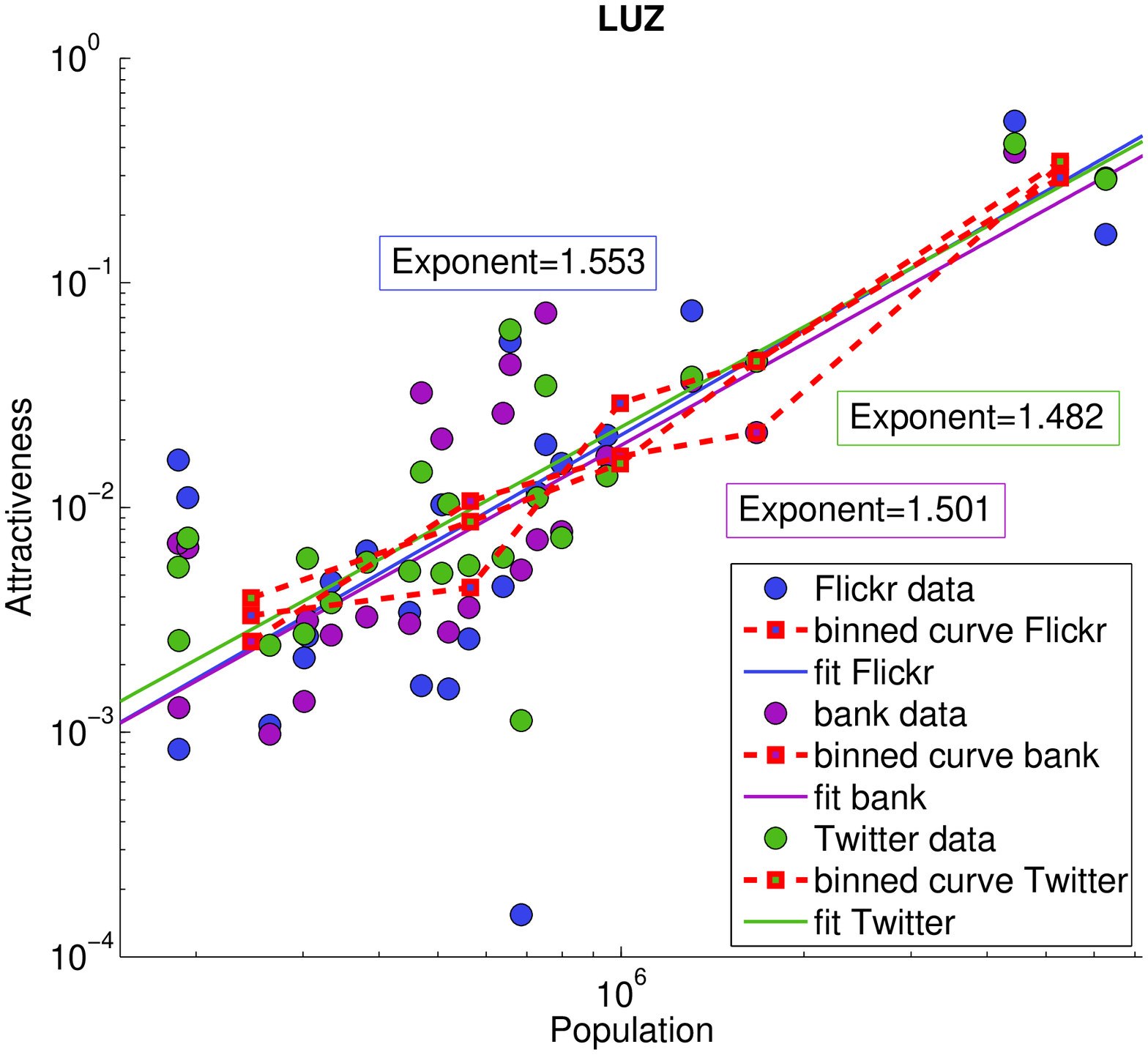}
\includegraphics[width=.49\textwidth]{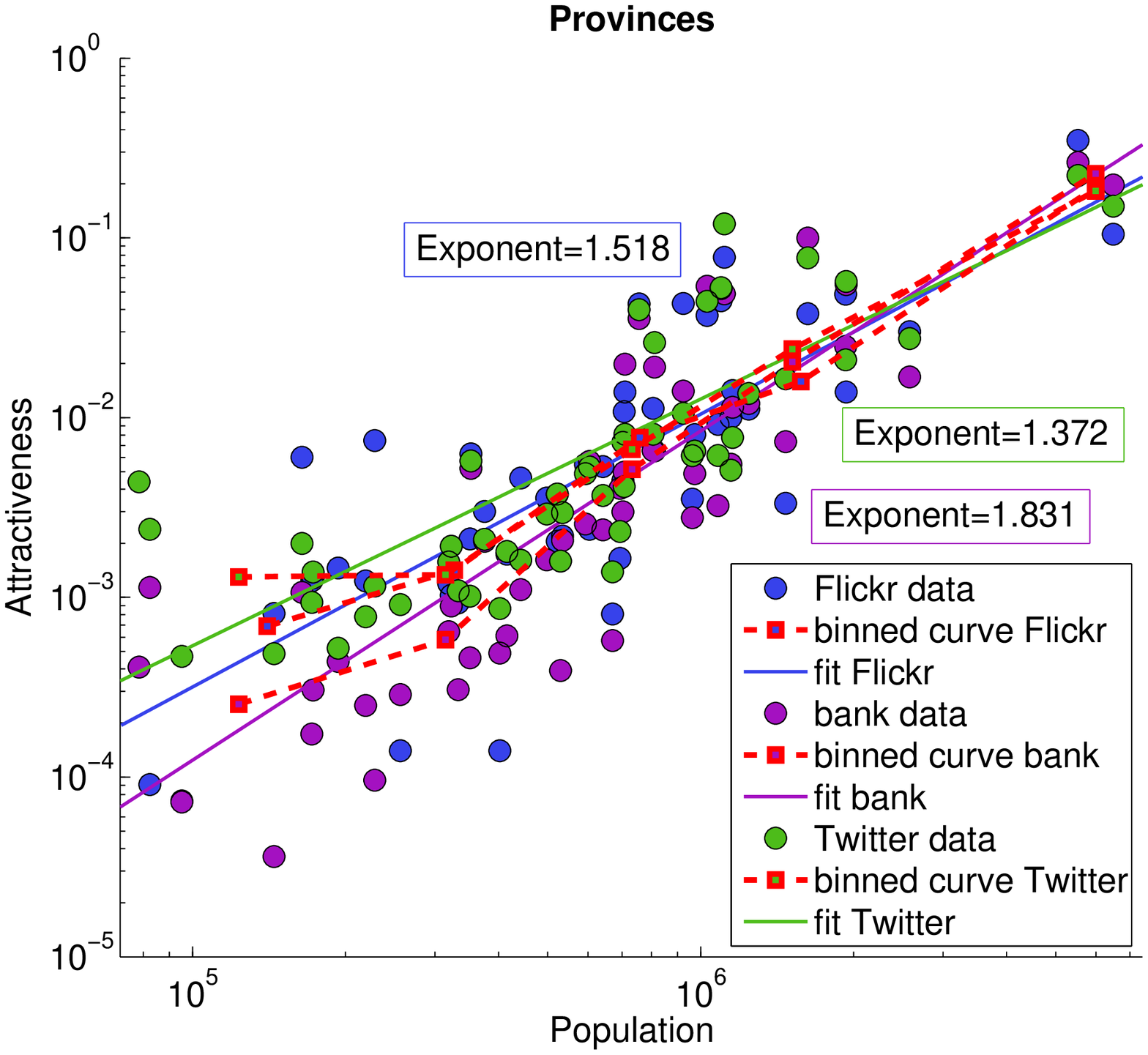}
\caption{\label{fig::city_scaling}Scaling of city attractiveness with population size observed through three different datasets for different Spanish city definitions as well as for the provinces. X-axis represents the number of people living in the city, while Y-axis represents the fraction of the number of photographs/videos, tweets or bank card transactions registered in the city versus the total amount registered in all the cities.}
\end{figure*}

Figure \ref{fig::city_scaling} reports the results of fitting a power-law dependence $A\sim a p^b$ to the observed pairs of attractiveness $A$ and the population $p$. Fitting is performed on the log-log scale where it becomes a simple linear regression problem $log(A)\sim log(a)+b\cdot log(p)$. The fitted scaling trends are substantially superlinear for for all three datasets - BBVA, Flickr and Twitter - and for all types of city definitions (i.e., CONs, FUAs, LUZs) as well as for the provinces. Regardless of the spatial scale used in the analysis, the scaling exponent $b$ remained approximately the same for all three datasets considered (i.e., around $1.5$ with the highest level of fluctuation for provinces, compered to the lowest one for LUZs) confirming the robustness of the pattern. Moreover, this pattern seems to be quite significant~-- such a high exponent indicates that for example attractiveness of one city that is $3$ times larger than the other one, is expected not to be $3$ times, but on average $5$ times higher.

In order to double-check if the average scaling trend is really consistent with the fitted power-law, we perform the binning of the data by considering average attractiveness of all the cities falling into each of the five population ranges, evenly splitting the entire sample on the log-scale. As one can see from \ref{fig::city_scaling}, the binned trend in all the cases goes pretty much along the fitted power-law, confirming the scaling shape.

Finally, the analysis of the fit statistics confirms statistical significance of the trends --- $R2$ values for the binned trends fitted by power laws vary as $97.6\pm 1.8\%$ for all three ways of city definition and all three datasets, quantitatively confirming the observed visual similarity between the trends and the superlinear power laws. Quantifying $R2$ for the power-law fits to the scattered plots including the entire variation of all the individual original city data still keeps $R2$ high enough, reporting $57.5\pm 12.9\%$ of the total data variation being explained by the superlinear trends. At the same time $p$-values are always below $1\%$ (usually much lower) in all the cases considered, serving as an ultimate evidence of the trend significance.

\subsection{Temporal aspects of city attractiveness}
In the analysis that was presented in Section \ref{section_3.1}, we considered the overall aggregated city attractiveness over the entire time frame of data availability (i.e., one year in case of Twitter and BBVA datasets or ten years in a case of Flicker data). However, one should be aware of the noticeable seasonality of visitation patterns. Therefore, in this section we investigate whether or not this seasonality affects the observed attractiveness scaling. For that purpose we consider an aggregation of the activity over a moving window of three-month seasons, shifting it month-by-month through the entire year.

\begin{figure*}[t!]
\centering
\includegraphics[width=.32\textwidth]{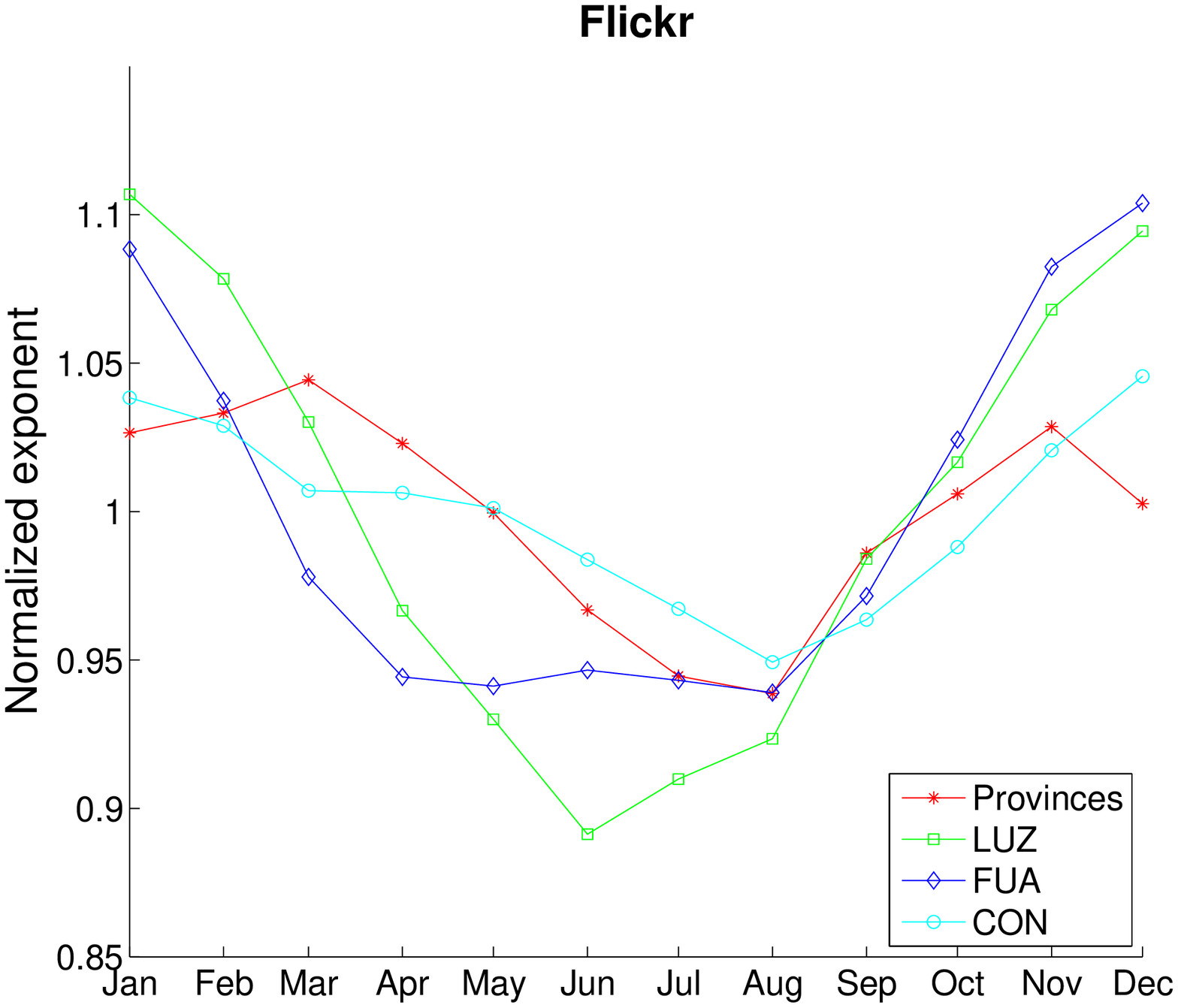}
\includegraphics[width=.32\textwidth]{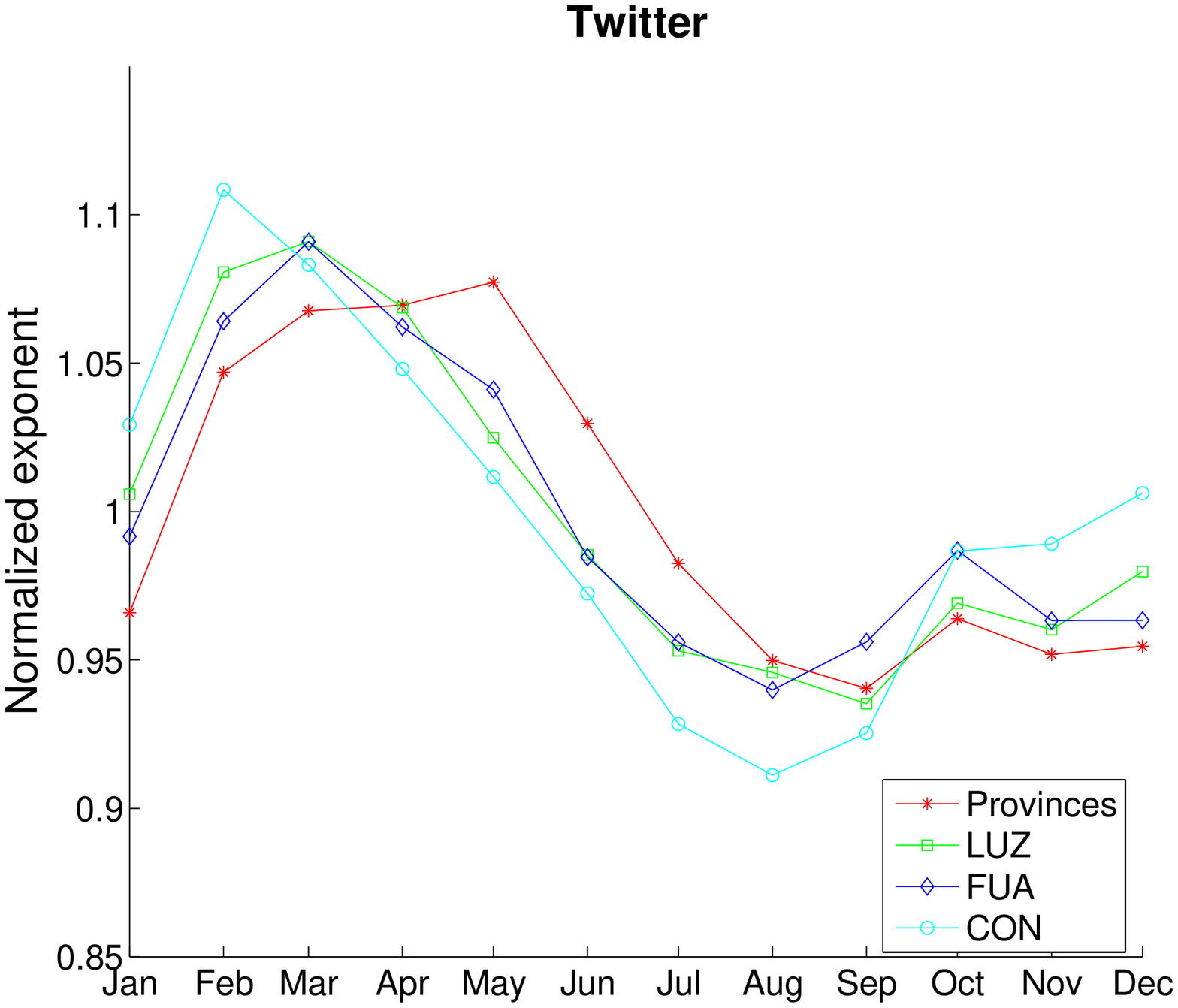}
\includegraphics[width=.32\textwidth]{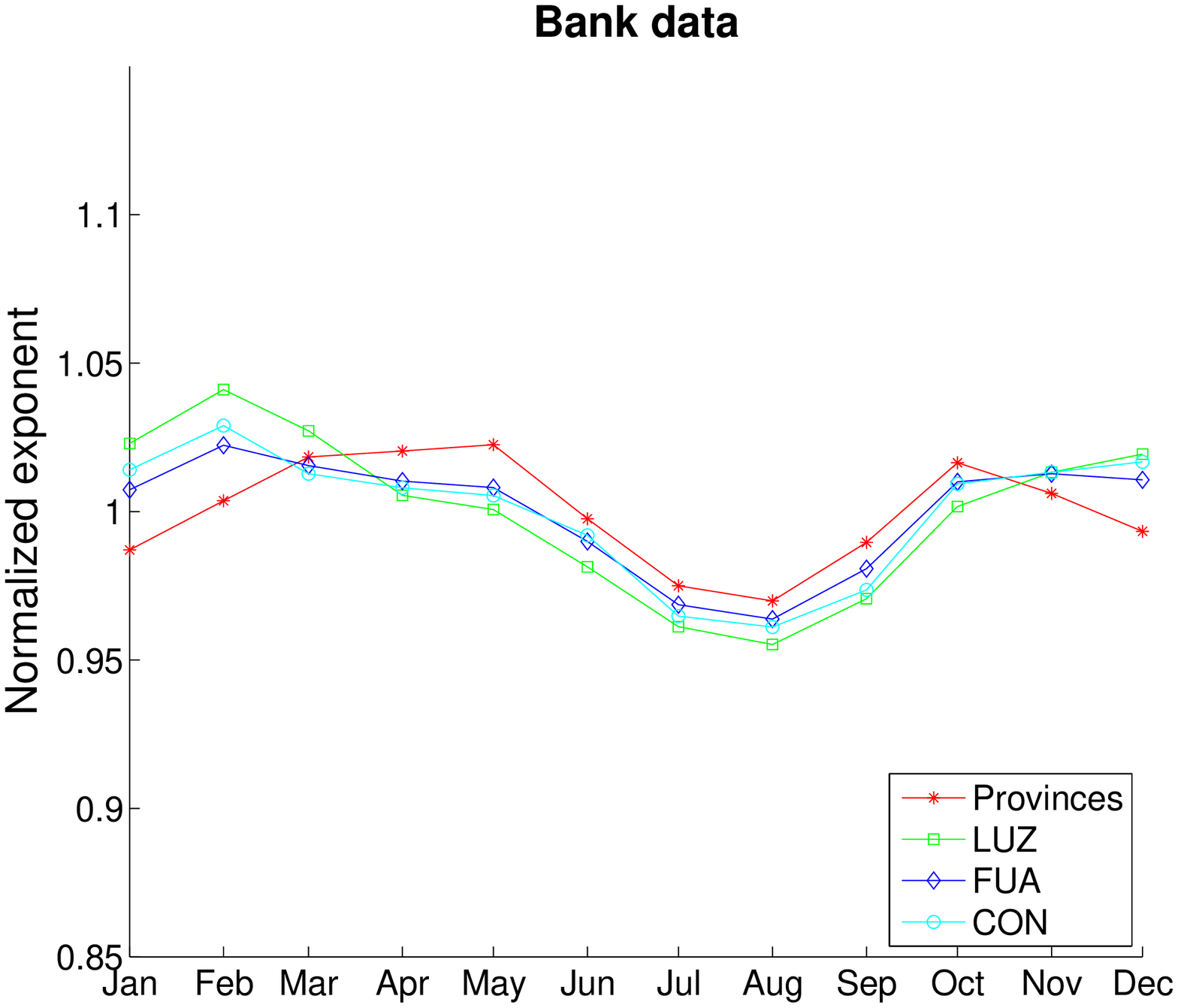}
\caption{\label{fig::exponent_by_monthes}Variation of the relative value of the scaling exponent over the year, normalized by the yearly average.}
\end{figure*}

\begin{figure*}[b!]
\centering
\includegraphics[width=.8\textwidth]{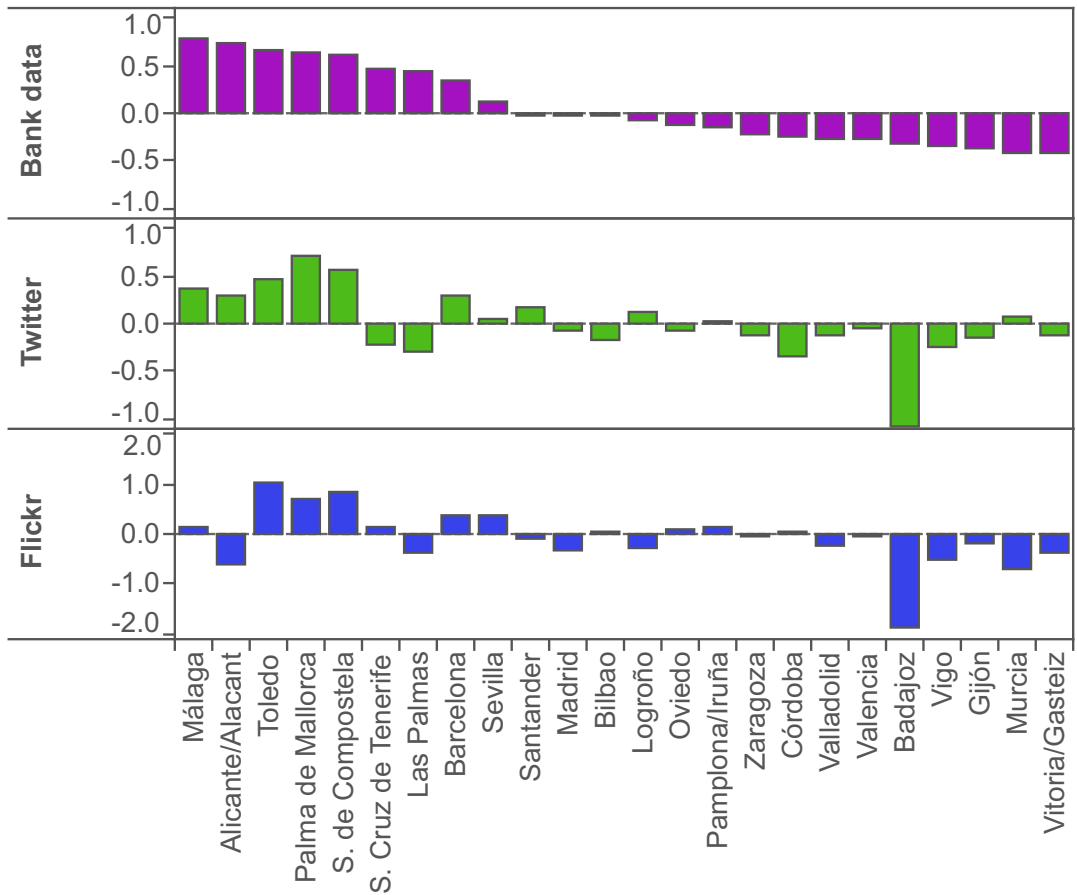}
\caption{\label{fig::fig_residuals}LUZ scale-independent attractiveness through three data sets.}
\end{figure*}

Figure~\ref{fig::exponent_by_monthes} shows a substantial dependence between the observed exponent and the time of the year (i.e., month). The trend appears to be mostly consistent across different ways of city definition, however shows slight variation depending on the dataset. However, the main pattern is always the same and is confirmed by all trends considered~-- it always drops over the summer as it seams that people tend to explore more extensively and more different destinations in Spain. This could be easily explained by a higher touristic activity over the summer, especially focused on beach tourism being spread over the number of smaller destinations along the coast, while the rest of the year, especially in spring and autumn, there is a larger number of business foreign visits who are primarily attracted to the major cities.

\subsection{Learning from deviations - scale-free city attractiveness}
The above superlinear power-law trends describe the way city attractiveness scales with the city size on average. However, each particular city performance can be different from the trend prediction. This actually opens up a possibility for the scale-independent scoring of the city attractiveness by considering the log-scale residual of the actual attractiveness value vs.\ the trend prediction: ${\rm res}=log(A)-b\cdot log(p) - log(a)$. This residual being positive points out to the city over-performance vs.\ the average trend, while the negative value points out that the city is under-performing.

Just to give an example, Figure \ref{fig::fig_residuals} visualizes residuals for the LUZs ordering the cities from the most over- to the most underperforming ones according to the bank card transactions data. One can notice that although residuals from different datasets are different, the patterns are generally consistent~-- cities strongly over/under-performing according to one dataset usually do the same according to the others. However, there are some interesting exceptions e.g., in cases of Malaga and Alicante. Although foreign people use to visit those cities, as one can see from bank card and twitter data, the visitors seem to have relatively lower motivation for posting pictures of them. 
This pattern could probably be explained - those places are known to be the typical retirement places for senior people from northern Europe \cite{munoz2012atractivo} - some of them move there, while many keep visiting these places especially over the winter. And it seems quite likely that visitors from this category might typically be much less active users of Flickr.
Similarly, the two island cities: Santa Cruz de Tenerife and Las Palmas are attracting foreign visitors to spend their money, but do not seem to encourage them enough to perform both online activities: Twitter and Flickr. Another interesting outlier is Badajoz which seems to be particularly underperforming in terms of Flickr and Twitter activity of the foreign visitors. A possible reason might be the context of this city not really being a major touristic attractor, but mainly, because of its proximity to the frontier, serving the nearby Portuguese people as a shopping center and service provider.

The outliers highlight the importance of a more in-depth analysis which could address not only the very fact of foreign visitor activity but also it's reasons as well as its structure by visitor origin and type of activity. Nevertheless, overall consistency of the residual values defined through different datasets also observed on the scales of FUAs, CONs and the provinces, highlights the robustness of the general patterns. Table \ref{tab:Correlations} presents the pairwise correlation values between those residuals which happen to be high enough, typically falling in the range between 50 and 80\%.
 
\begin{table}[h]
\caption{\label{tab:Correlations}Correlations (\%) between city/provinces residuals defined through different datasets.}
\begin{center}
\begin{tabular}{|c|c|c|c|}
\hline
            & bank/twitter  & bank/flickr   & twitter/flickr \\
\hline
Provinces   & 84.45        & 46.77        & 52.89 \\
LUZs        & 62.95        & 52.72        & 77.09 \\
FUAs        & 73.68        & 59.98        & 77.90 \\
CONs        & 80.07        & 56.90        & 58.56 \\
\hline
\end{tabular}
\end{center}
\end{table}

%\newpage

\section*{Conclusions}
\label{conclusion}

In this study we leveraged three types of big data created by human activity for quantifying the ability of cities in Spain to attract the foreign visitors. In general, city attractivity was found to demonstrate a strong superlinear scaling with the city size. A high consistency of the scaling exponents across different ways of defining cities as well as across all three datasets used in the study provides an evidence for the robustness of our finding and also serves as an indirect proof of the applicability of selected datasets for that purpose.

Moreover, we analyzed the temporal variation of the scaling exponent during a year, which was found to reveal a very intuitive pattern quantified by a noticeable drop of the exponent value over the summer~-- visitor activity seems to be more spread across different smaller destinations within the country over more touristic summer time, while more concentrated at major destinations in autumn and spring, presumably because of having more business visitors in the country. Again the pattern appears to be pretty robust and consistent as being confirmed by all the datasets for all different city definitions.

\section*{Acknowledgments}
The authors would like to thank Banco Bilbao Vizcaya Argentaria (BBVA) for providing the anonymized bank dataset as well as Eric Fisher and Yahoo! Webscope program for providing the Flickr datasets for this research. Special thanks to Assaf Biderman, Marco Bressan, Elena Alfaro Martinez and Mar\'ia Hern\'andez Rubio for organizational support of the project and stimulating discussions. We further thank BBVA, MIT SMART Program, Center for Complex Engineering Systems (CCES) at KACST and MIT, Accenture, Air Liquide, The Coca Cola Company, Emirates Integrated Telecommunications Company, The ENEL foundation, Ericsson, Expo 2015, Ferrovial, Liberty Mutual, The Regional Municipality of Wood Buffalo, Volkswagen Electronics Research Lab, and all the members of the MIT Senseable City Lab Consortium for supporting the research. Part of this research was also funded by the Austrian Science Fund (FWF) through the Doctoral College GIScience (DK W 1237-N23), Department of Geoinformatics - Z\_GIS, University of Salzburg, Austria. 

%\newpage

\bibliography{literature}

\end{document}